# First experimental demonstration of an isotropic electromagnetic cloak with strict conformal mapping

Yungui Ma[1, ‡], Yichao Liu[1,2,‡], Lu Lan[1], Tiantian Wu[1], Wei Jiang[1], C. K. Ong[2] & Sailing He[1,3,*]

[1]*Centre for Optical and Electromagnetic Research, State Key Lab of Modern Optical Instrumentation, Zhejiang University, Hangzhou 310058, China*
[2]*Department of Physics, National University of Singapore, 2 Science Drive 3, Singapore 117542*
[3]*Department of Electromagnetic Engineering, School of Electrical Engineering, Royal Institute of Technology, S-100 44 Stockholm, Sweden*

**In the past years quasi-conformal mapping has been typically used to design broadband electromagnetic cloaks. However, this technique has some inherit practical limitations such as the lateral beam shift, rendering the device visible or difficult to hide a large object. In this work we circumvent these issues by using strict conformal mapping to build the first isotropic cloak. Microwave near-field measurement shows that our device (with dielectric constant larger than one everywhere) has a very good cloaking performance and a broad frequency response. The present dielectric approach could be technically extended to the fabrication of other conformal devices at higher frequencies.**

[‡]These authors contributed equally to this work.

*Correspondence to sailing@zju.edu.cn

Modern development of transformation optics (TO) has enabled the design and demonstration of many fancy electromagnetic devices[1-8]. Among them, a successful application is to provide concrete blueprints for invisible cloaks[1-3]. However, the bright future of this technique is doomed by the complexity of the transformed media that are very difficult for practical applications. For a cloaking device, the primary problem is how to effectively fulfill strong inhomogeneous anisotropic material parameters typically with sub-unity requirements. Since Pendry's first cloaking theory[1] and successive experiment[3], the EM cloak has experienced important advancements both in frequency bands from microwave to optics[9-12] and in geometries evolving from two to three dimensions[13-17] excited by the modification of the original theory using quasi-conformal mapping[18]. However, as we know now this technique involves unequal coordinate-axis compression/extension in transformation and only produces weak anisotropic cloaking media[19]. One of the disadvantages is that it inherently makes a device actually detectable parallel to the lateral direction[19,20]. The other problem is in implementation as it is very hard to configure a device to hide large objects under the isotropic material approximation[21]. This can be the same reason that limits the usage of weakly anisotropic natural crystals such as calcite to achieve relatively thin cloaks compared with the invisible region, as recently demonstrated by several groups[22-24], which were theoretically inspired by the bilinear general transformation [25, 26].

So far the quasi-conformal carpet-type cloak has been most explored in the literature for a half-space ground hiding. In 2009 Cui et al first experimentally used this technique to acquire a full-space cloak by mirroring the invisible carpet together with the metal ground along the bottom axis. In this situation an object inside the cavity enclosed by the original and mirrored grounds was invisible for a far-field observer thus rendering a unidirectional cloaking effect [27]. In a later theoretical paper they proposed a line-transformed technique to have better unidirectional cloaking performance using the general transformation to avoid the inherit limits of a quasi-conformal cloak at the price of introducing anisotropic materials [28]. Recently Smith at el demonstrated a similar unidirectional cloaking device using homogeneous anisotropic metamaterials inspired by the bilinear transformation [29]. Their device elaborated good cloaking behaviors for only transverse electric (TE) waves and also only within a limited bandwidth because they have to use magnetic resonant elements and corrugations to realize the anisotropic sub-unity permeability profile. To the best of our knowledge, so far there is still no experimental report totally solving the application issues such as the lateral beam shift, cloak size and bandwidth for carpet or unidirectional cloaking.

In this work we overcome these critical problems by first using strict conformal mapping in experiment to achieve an isotropic dielectric invisibility device and demonstrate both unidirectional and carpet-type cloaking behaviors in a wide band of microwave frequencies. Our design theory is equivalent to the one described in Leonhardt's 2006 paper on non-Euclidean transformation for cloaking[2]. However, here we only consider the first Riemann virtual sheet and reorganize the spatial geometry to obtain a unidirectional cloak for a full physical space. By adopting the half dielectric profile we fabricate a second device to demonstrate the carpet-type cloaking. Application of the conformal mapping guarantees an isotropic dielectric device without any issue of e.g. lateral beam shift[30]. The mapping relationship and geometrical truncation can be properly conceived to control the thickness of the final cloak. These measures help us to avoid the drawbacks encountered by the previous methods as analyzed above and show wideband cloaking performance. The dielectric approach developed in this paper can be transferred to higher frequencies to enable the manufacture of other conformal devices by advanced fabrication technologies such as three-dimensional (3D) printing or lithography.

**Results**

In the reference[2], conformal mapping was first introduced to achieve a full-space invisible cloak under the condition of geometrical optics where the existence of a second Riemann sheet plays a key role in rendering a hiding space. In the process the Zhukovsky transform is used to configure the coordinate projection between the virtual ($w$) and physical ($z$) spaces with a simple formula

$$w = z + \frac{a^2}{z} \quad (1)$$

where $a$ is a key parameter defining the final cloak size. However, the transformed material for the second virtual sheet is located inside a limited physical volume with complicated parameters and is very difficult to implement [2]. Here, instead of realizing a perfect cloak, we use this mapping formula to generate an isotropic line-transformed cloak by sealing off the branch-cut (entrance to the second Riemann sheet) with a metal wall as shown by the yellow line in Fig. 1a[31,32]. From Eq. (1), we know this operation will blow up an invisible black hole in the transformed physical space, as shown in Fig. 1b, and the flat coordinate of the original Cartesian space in Fig. 1a will be crushed to form an infinitely large cloak shell. The red and blue lines in the figures represent light rays illuminating the sample with directions parallel and perpendicular to the metal walls (yellow lines), respectively. It is obvious that the red light will pass through the cloak and restore their directions in the far field without disturbance while the blue light will be mirrored back when it touches the metal wall. This conformal transformation configures an ideal unidirectional cloak with an effective hiding region of diameter equal to $a$. A far-field observer will view a piece of metal sheet (or mirror in optics) of length = $4a$ from the off-axis angles.

The material parameters of the device are calculated by $n = \left|\frac{dw}{dz}\right| n'$ with $n'$ being the refractive index in the virtual space.

**Implement the unidirectional cloak**

Now we go to the fabrication step. It will not be easy to manufacture such an isotropic cloaking device discussed here because originally this device is infinitely large in size and has a local index distribution close to zero around the two ends at $z = \pm a$. The low rotational index symmetry is another difficult aspect in the experiment. Here we overcome these difficulties by taking three approximations: (i) The device size is suppressed by truncating the outer part where local indices are almost of no variation and the rest as a cloak device has a reasonable size compared with the cloaked region to achieve a small cloak device. Here we choose the cloak boundary (indicated by the dashed magenta circle in Fig. 1b) at radius $r = 5a$ (four times larger than the invisible hole). (ii) We assume the background of the virtual space has a moderate index $n' = 2$, and thus the region of sub-unity index in the transformed device is very small and can be approximated by air in the final fabricated device. (iii) The continuous cloaking model is approximated by numerous homogeneous elements with a dimension of about one tenth of local wavelength.

The validity of the above approximations is numerically examined by simulating the wave scattering for different samples as plotted in Figs. 2a to 2c. Here the incident Gaussian beam has a waist of 6 cm and illuminates the samples from the left side with the direction parallel to the red line in Fig. 1b. The example frequency is 5.5 GHz and the cloak parameter $a$ equals to 3 cm. Perfect magnetic conducting (PMC) conditions are utilized to define the outer boundaries of the cloaked region. This selection is technically required because only a magnetic conducting line segment obeys the free-space symmetry for the excitation of the TE wave [29]. Figure 2a gives the electric field pattern for the ideal case without any approximation. It shows a perfect unidirectional cloaking behavior. The reduction of the index distribution by replacing the sub-unity region with air, as shown in Fig. 2b, does slightly degrade the overall unidirectional cloaking performance with irregular side scattering. However, the good cloaking effect is basically retained. Figure 2c shows that the successive operations of truncation and discretization applied to the device have little effect on the cloaking performance. These results validate our parametric and structural approximations utilized in the experiment discussed below.

To implement the truncated device, we divided the whole 12-cm-thick cloak into 31 layers and 60 angular sectors. Each unit element has a dimension of about one tenth of the local wavelength as the permittivity ($\varepsilon$) varies from 1 to 14.5. Rectangular metallic inclusions of different sizes have been fabricated using standard lithography of printed circuit boards (Rogers 3850) to realize the gradient index profile as shown in Fig. 3a. Dielectric powders of different dielectric constants have been used to fulfill the requirement of the wide range index. Great care has been taken to correctly fill the right powders into the right small meshes, which consumed most of the fabrication time. In order to implement the PMC boundary as required by TE polarization, here we separated the hiding region of PEC cavity from the cloak by a quarter wavelength dielectric doped with magnetic material (ECCOSORB® SF10 rub absorbers) [5,33]. The later experiment shows it has a good wideband PMC approximation. Figure 3b gives a photo of our implemented device. It has a full diameter of 300 mm and height of 6.4 mm. Background material consisting of powders of $\varepsilon$ = 4 has been used to match the impedance at the device boundary. The rectangular edge portions of the background have also been linearly trimmed to produce a gradient height change in order to reduce the edge scattering. The red arrow in Fig. 3b represents the incident beam direction parallel to the red lines in Fig. 1b. More details about fabrication and measurement can be found in the method section.

The measured field patterns are shown in Figs. 4a to 4d at the measuring frequencies of 4.0, 5.0, 6.0 and 7.0 GHz. From these figures we can see that the unidirectional cloaking effect is well achieved by our sample. The animation of the wave propagation at the example frequency of 5.0 GHz is given in Fig. S1 in the online supporting document. The slightly broadening of the wave beam in Figs. 4a to 4c after passing the hiding region may be caused by two main reasons: one is the deviation of the local index from the theoretical value due to the imperfect manufacture and the other is the curved incident wavefront produced in experiment rather than the strict planar beam used in simulation. Our device was designed to work at frequencies below 6 GHz, beyond which the dispersions of the embedded metamaterial elements (metallic inclusions) become severe. This point is manifested by the deteriorated cloaking performance as shown in Fig. 4d for the frequency of 7.0 GHz. In principle the bandwidth problem could be solved using pure natural dielectric materials to build the device enabled by our design.

**Implement the carpet-type cloak**

The carpet-type cloak predominantly discussed in the literature[9-17,31,34-38] is a special case of our current design. It is fabricated by using a half dielectric profile with a PEC bottom boundary. The simulation and experimental results are given in Figs. 5a to 5f at 5.5 GHz. As a comparison we first show a numerical reflection pattern of a flat mirror in Fig. 5a. This effect is basically reproduced by our device model as shown in Fig. 5b for continuous media and Fig. 5c for discrete media. The discretization does slightly increase the irregular scattering but is still acceptable at our designed working frequency below 6 GHz. Close inspection on the power flow lines shows that the reflected wave beam has negligible lateral shift consistent with the theory, while lateral beam shift is usually observed in a quasi-conformal carpet cloak. Figure 5d gives a picture of our carpet-type cloak. The red arrow indicates the incoming wave direction and the black line represents the metal ground. The measured field pattern shown in Fig. 5e exhibits a good mirror reflection effect and agrees with the numerical result in Fig. 5f (a zoom-in figure from Fig. 5c). The animation wave pattern showing this behavior is given in figure S2 in the supporting document. The measurement shows a similar carpet-type cloaking effect at frequencies below 6 GHz, which is consistent with the previous experimental results for the unidirectional cloaking.

**Discussions**

We have utilized strict conformal mapping to design and fabricate an isotropic cloaking device by avoiding the drawbacks of the previous samples designed by the quasi-conformal technique. The measurement results verify the validity of our approximations made in the fabrication and elaborate good unidirectional and carpet-type cloaking behaviors. Another important result of this experiment is to show that conformal devices could be well manufactured to exhibit outstanding performance comparable with ones designed by the general TO method but with the additional advantages such as improved bandwidth response and fabrication feasibility. With the help of advanced technologies such as 3D printing or lithography, it can be highly anticipated that similar conformal dielectric devices could be fabricated at THz or even optical frequencies with the index profiles carefully handled.

Here we would like to emphasize that transformation optics and the cloaking devices have already been well understood for many years and currently the key thing is how to correctly implement them. This is in line with the recently published unidirectional cloaking experiment working on TE polarization although the underlying transformation theory is well known before [25,26]. The previous isotropic cloaks designed by the quasi-conformal mapping have relatively small permittivity distribution (e. g., $\varepsilon$ = 1.17 to 2.79 for the first microwave carpet[10]) and thus are much easier for fabrication by just using patterned printed circuit boards or dielectric hole arrays[15]. In this work, we use a strict conformal mapping method to generate a relatively larger invisible region enabled by a larger index distribution for the cloak, i.e., $\varepsilon$ = 1 to 14.5. Thanks to the different dielectric powders available (one may use e.g. Germanium for optical frequencies), we eventually managed to fabricate such a device and experimentally verified the cloaking performance. This helps to demonstrate the technical possibility of a rather complicated isotropic device. The quality of our device and the eventual performance could be enhanced by using advanced fabrication techniques as mentioned above. We believe the technical methods developed in this paper are very valuable for the development of other interesting isotropic devices designed by the conformal technique, such as the recently reported conformal lenses having illusionary optical effect[32] or perfect imaging lenses[33].

**Methods**

The simulation in this work was done by the commercial software COMSOL. The largest mesh size is smaller than one tenth of local wavelength. We use a Gaussian beam (waist = 6 cm) to illuminate the device along two directions, i.e., parallel and orthogonal to the branch-cut line as shown in Fig. 1a. In the simulation and later experiment, the radius of the invisible hole is 30 mm and the outer radius of the cloak is 150 mm. This corresponds to an invisible-region-to-cloak thickness ratio of 1/4, which is already larger than the previous samples designed by the quasi-conformal mapping. To implement the device, we divided the 120 mm thick cloak shell into 31 unequal layers along the radius direction and 60 sectors along the angular direction. Metamaterials made of rectangular metallic inclusions by lithography of Rogers 3850 circuit boards have been designed and used to realize the spatial index distribution. Emerson dielectric powders of epsilon = 2.5, 5 and 7 have been used together with air as background filling materials. The implemented device is 300 mm in diameter and 6.4 mm in height. Along the height there are two layers of metamaterial elements. To mimic a PMC boundary to separate the cloak from the invisible region, we utilized two layers of Emerson SF10 microwave absorber sheets (totally in 3 mm thickness) backed on the surface of an Al cylinder. This microwave absorber acting as a quarter wavelength spacer can reduce the electric field scattering around the inner metals. Our cloak device is embedded in a dielectric background of epsilon = 4. We used the trimmed plastic form bars to create matched background edges by inducing a gradient height variation near each side of the rectangular background edges. At about a distance of 3 cm, we placed some microwave absorption foams to reduce the internal scattering. For the carpet-type cloak, we use metal walls as the bottom ground and on the top use the embedding powders to enclose two orthogonal edges. The wave is illuminated from the top-left edge and reflected out through the top-right edge. In measurement, the sample was placed inside a parallel-plate waveguide chamber mounted on a platform stage driven by a *xy*-step motor. A vector network analyzer (ZVA 40) was used to generate and process the microwave signals which were guided into the measurement chamber through two coaxial cables (feeding port: a 120 mm x 6.4 mm rectangular waveguide; detection port: a 2 mm thick monopole antenna). The minimum scanning step size (or the resolution of the field picture) is 1 mm. The time interval between two points is controlled to be no less than one second (slow scanning is critically required to reduce the displacement of the loose powders).

**Acknowledgments**

The authors are grateful to the partial supports from NSFCs 61271085, 60990322 and 91130004, the National High Technology Research and Development Program (863 Program) of China (No. 2012AA030402), NSF of Zhejiang Province (LY12F05005), the Program of Zhejiang Leading Team of Science and Technology Innovation, NCET, MOE SRFDP of China, the DIRP grant of Singapore (R144000304232), and Swedish VR grant (# 621-2011-4620) and AOARD.

**Author Contribution**

M.Y.G. conceived the ideal and L.Y.C. designed the experiment. L.Y.C, M.Y.G., L.L., W.T.T. and J.W. participated in the fabrication and measurement of the samples. M.Y.G. and S.H. wrote the article and all the other authors reviewed it. S. H. supervised this study and finalized the manuscript.

**Additional information**

Competing financial interests: There is no competing financial interest for this work.

**Captions**

Fig. 1 Conformal transformation scheme. (a) Cartesian virtual space and (b) transformed physical space. The yellow metal wall of length $4a$ in (a) is expanded into a circle in (b) enclosing a dark invisible region of radius $a$. The red and blue lines represent two types of wave beams of orthogonal incident directions. The blue light ray hitting the metal wall will be mirrored back and the others will travel forward without disturbance. The magenta dashed line in (b) describes the outer profile of the truncated cloaking device.

Fig. 2 Simulated cloaking behaviors. (a) The ideal case with no parametric and structural approximations; (b) the case with reduced index distribution where sub-unity regions are replaced by air; (c) the case of a truncated and discretized cloak device. PMC boundaries are used here to isolate the invisible region from the outside EM wave.

Fig. 3 Unidirectional cloak. (a) Permittivity profile for the discrete device implemented here and (b) a photograph of the implemented device placed on the measurement platform. The relative permittivity has a distribution from 1 to 14.3, which changes dramatically around the invisible hole. Filling powders of different dielectric constants are used here. The white lines in (a) profile the discrete elements implemented. The red arrow in (b) represents the incident wave direction parallel to the red lines in Fig. 1b.

Fig. 4 Near-field measurement results for unidirectional cloak. (a), (b), (c) and (d) correspond to the measuring frequencies of 4.0, 5.0, 6.0 and 7.0 GHz, respectively. The cloaking performance is degraded at higher frequencies because of the dispersions of the metamaterial inclusions used in our samples. The cyan disk represents the hiding region.

Fig. 5 Carpet-type cloak. (a) Simulated reflection from a flat mirror, (b)/(c) simulated wave scattering for our carpet device with a continuous/discrete index profile respectively, (d) a photo of the carpet device, and (e)/(f) measured/simulated pattern of scattered field, respectively. Wave beam is illuminated from the top-left side at 5.5 GHz, as indicated by the red arrow in (d) which has a 45°angle deviation from the normal of the yellow branch cut in Fig. 1a. Similar cloaking effect is observed at frequencies below 6 GHz.

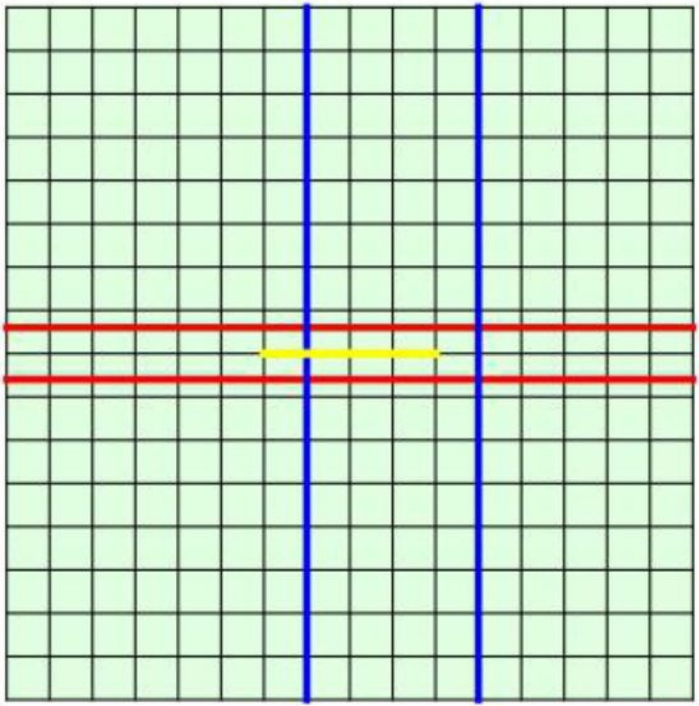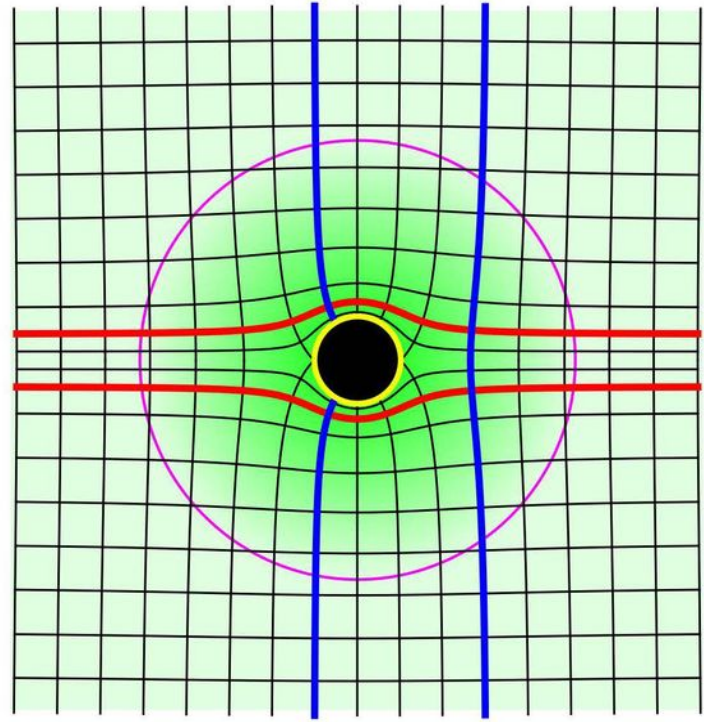

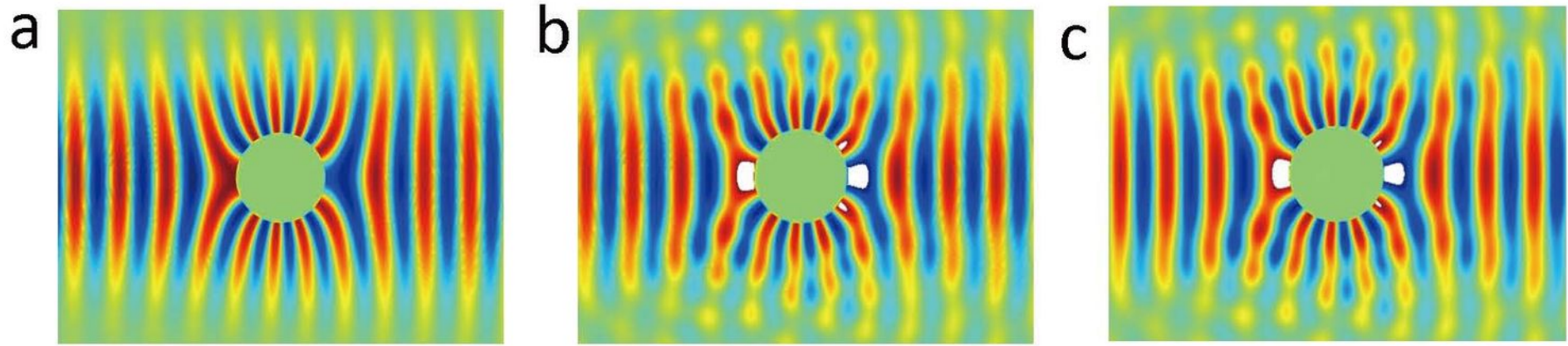

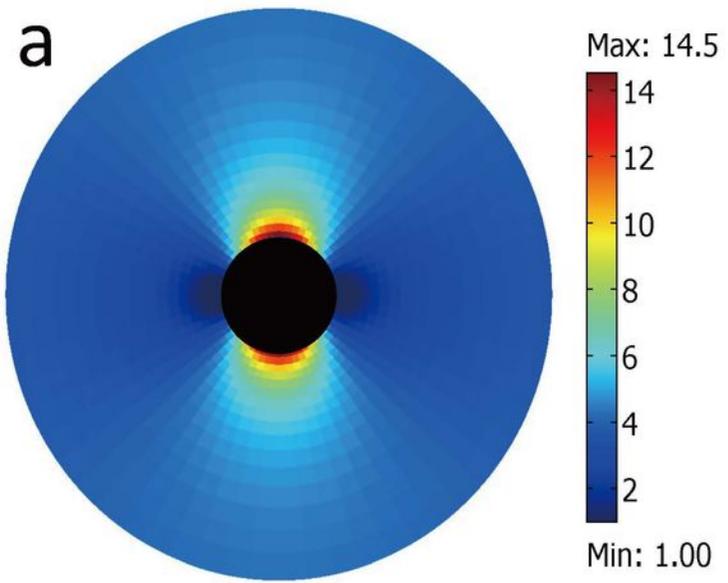 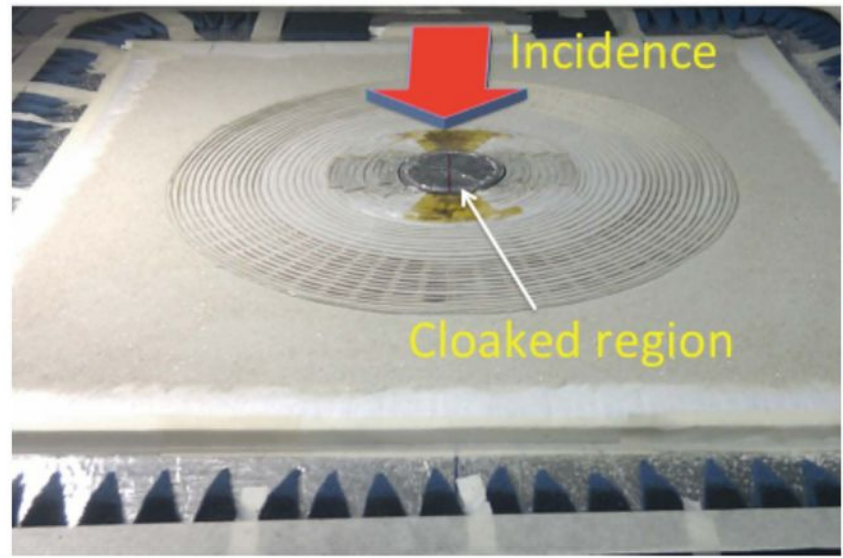

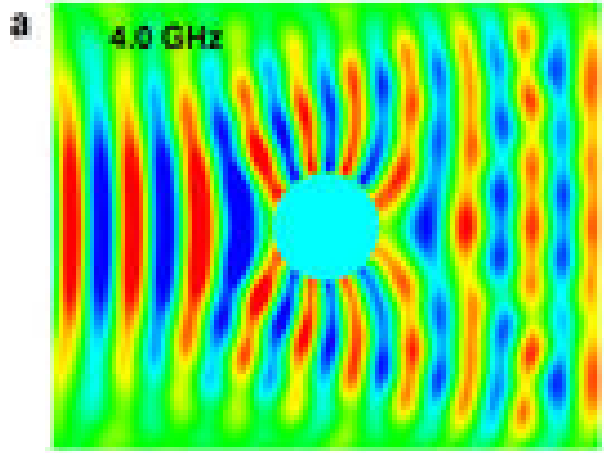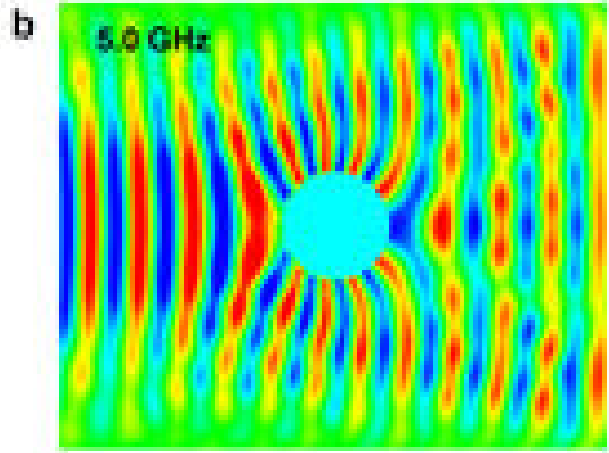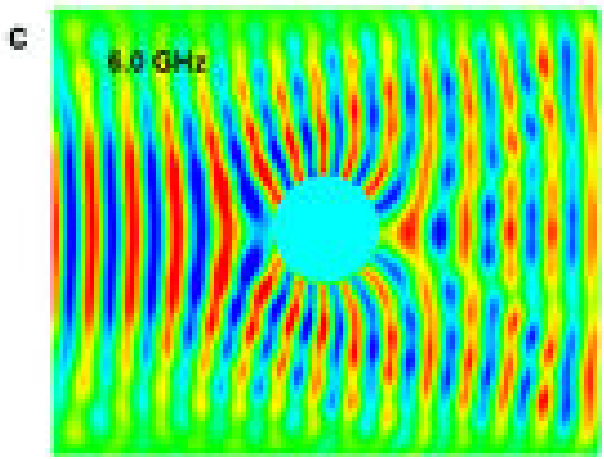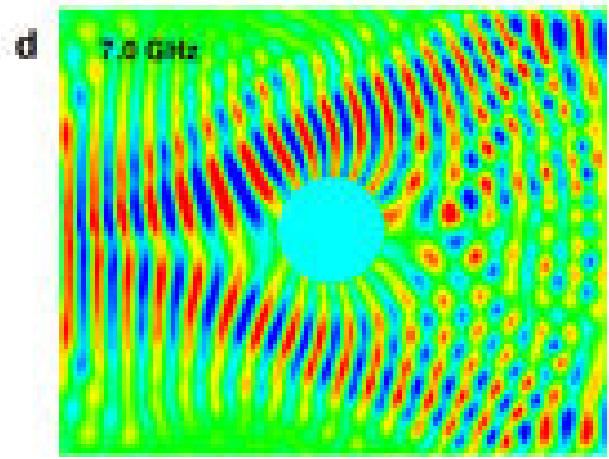

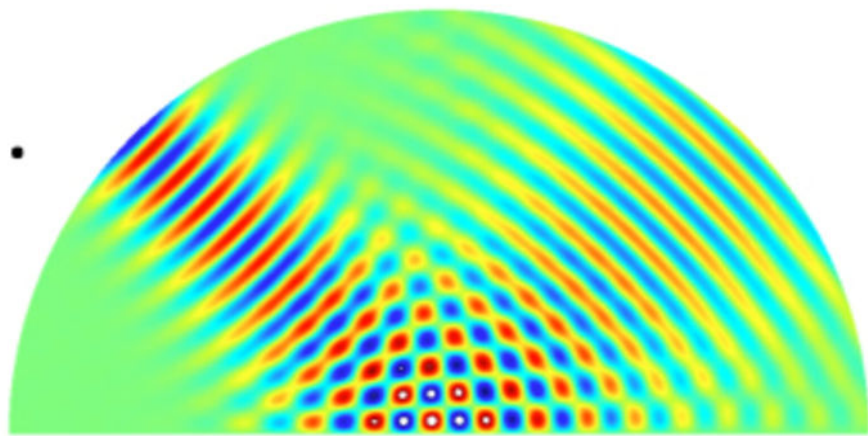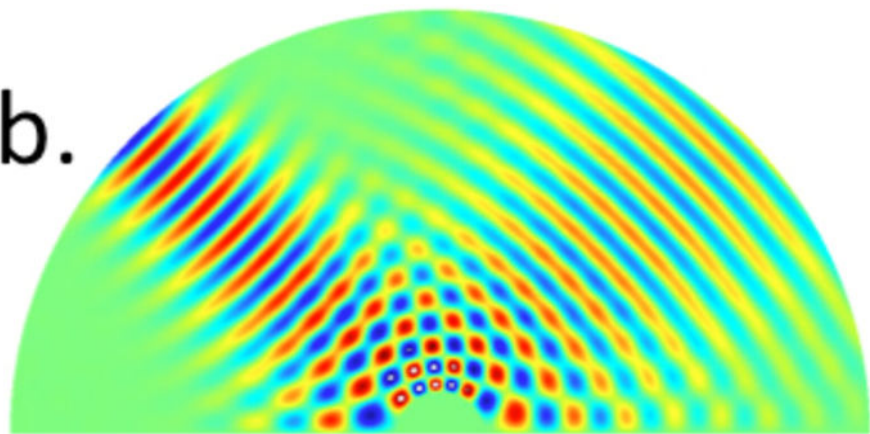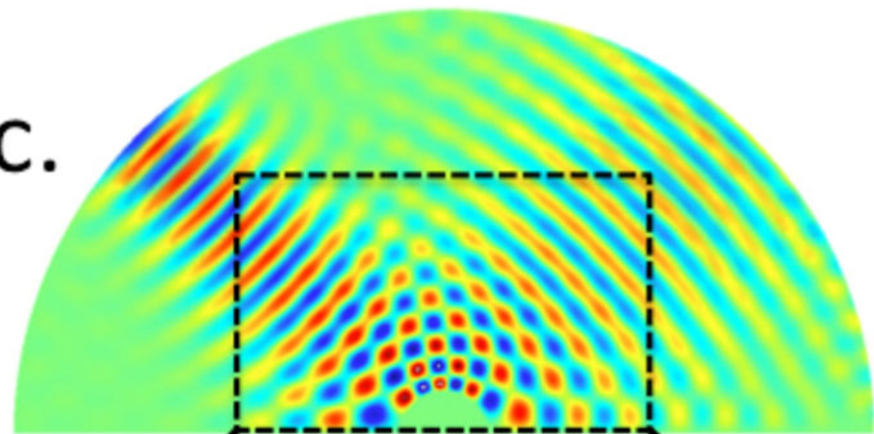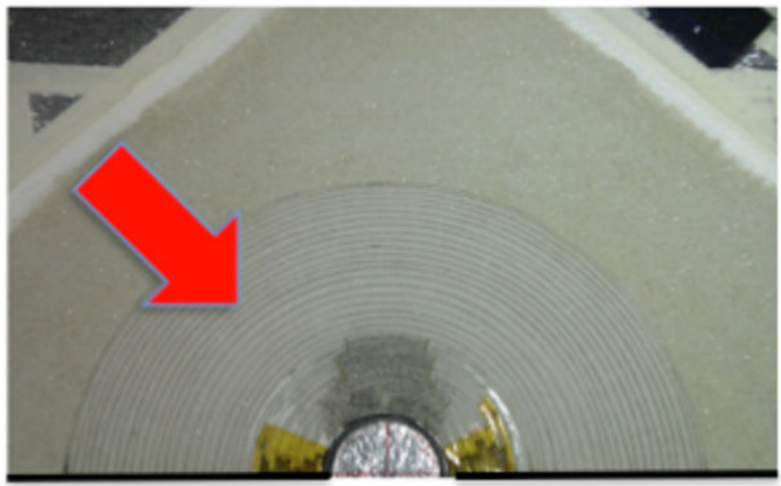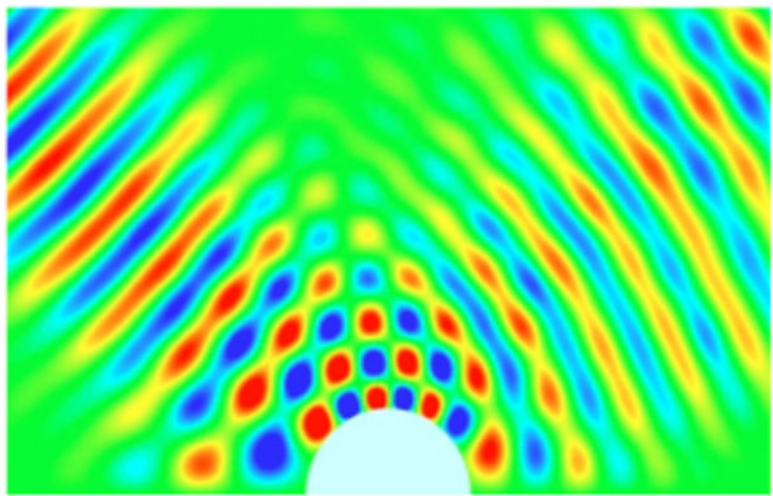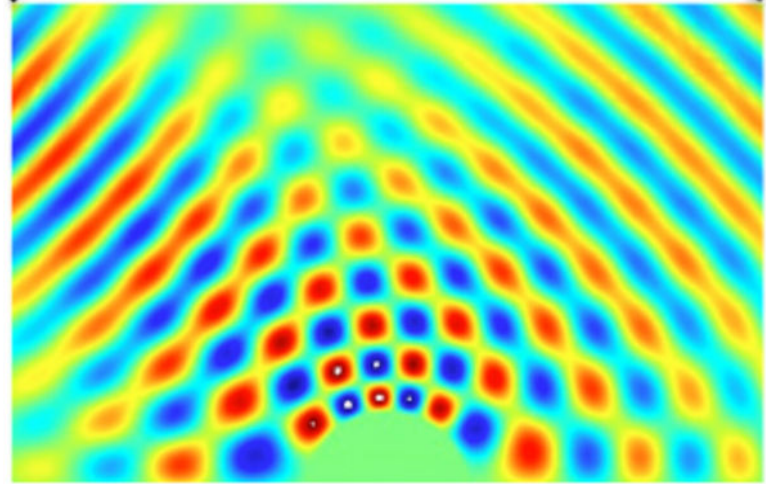